# Reconstruction of C&C Channel for P2P Botnet

Mohammad Jafari Dehkordi [1], Babak Sadeghiyan [1*]

[1] Computer Engineering and Information Technology Dept., Amirkabir University of Technology, Tehran, Iran
[*] basadegh@aut.ac.ir

**Abstract:** Breaking down botnets have always been a big challenge. The robustness of C&C channels is increased, and the detection of botmaster is harder in P2P botnets. In this paper, we propose a probabilistic method to reconstruct the topologies of the C&C channel for P2P botnets. Due to the geographic dispersion of P2P botnet members, it is not possible to supervise all members, and there does not exist all necessary data for applying other graph reconstruction methods. So far, no general method has been introduced to reconstruct C&C channel topology for all type of P2P botnet.

In our method, the probability of connections between bots is estimated by using the inaccurate receiving times of several cascades, network model parameters of C&C channel, and end-to-end delay distribution of the Internet. The receiving times can be collected by observing the external reaction of bots to commands. The results of our simulations show that more than 90% of the edges in a 1000-member network with node degree mean 50, have been accurately estimated by collecting the inaccurate receiving times of 22 cascades. In case the receiving times of just half of the bots are collected, this accuracy of estimation is obtained by using 95 cascades.

## 1. Introduction

With the spread of Internet access and their widespread use among people, the number of malicious users has also increased. Malicious users have always been seeking to reach their goals by creating and distributing malware in computer networks and infecting machines connected to these networks. In order to hide and gain more power, malicious users tend to use the resources of infected hosts. In recent years, dealing with organized malware has gained more importance.

The botnet is an important category of organized malware. Botmasters remotely control and lead infected hosts by issuing commands to those bots which can find access through command and control (C&C) channels. Investigators seek to unveil hidden dimensions of a crime by analyzing evidence gathered from C&C channels; however, botnet developers tend to adopt diverse techniques in order to hide the evidence.

Even though researchers try to detect a botnet and uncover its function details through botnet analysis, botnet developers make efforts to prevent easy analysis by using anti-forensics techniques such as encrypting and packing. Moreover, botnet developers hide the origin of the issued commands through changing the architecture of C&C channel from centralized to decentralized such as peer-to-peer (P2P) network.

In a P2P botnet, each command is forwarded by botnet members in several steps, delivering the command to all members. In this propagation, not all botnet members have direct connections with botmaster and this makes the investigator unable to discover the botmaster and break down the C&C channel. Being aware of the network topology for a P2P botnet (i.e., C&C channel) is a great asset for investigators to resolve questions related to P2P botnet forensics [1]. This knowledge facilitates the discovery of botmaster and the development of countermeasures against botnet attacks.

Most of the previous studies on P2P botnets have focused on the issue of detection, and finding botnet topology has not been addressed. In studies proposing botnet countermeasures and methods for finding botmaster, the approaches are not based on the reconstruction of the C&C channel as well [2]. While finding this topology, can help to resolve related issues more effectively and precisely.

In this paper, we propose a method for the reconstruction of P2P botnet C&C channel topology, which can accurately determine the connections among bots. The focus of this paper is on botnets, whose P2P network is based on a random model.

In our method, the topology of the botnet is reconstructed by employing a probabilistic approach. Moreover, in this method, we use the parameters of botnets graph model, the inaccurate receiving time of data in some of the botnet members in certain cascades, and end-to-end delay distribution on the Internet. Since the necessary data for our method does not depend at all on the type of P2P botnet protocol and the content of the message, our method can be used for all botnets.

The parameters of the botnet graph model include the number of nodes and their degrees mean. The P2P botnet members can be detected by methods such as enumeration or monitoring of the reactions to original or forged commands. The degrees mean can be measured by analyzing bot code or some node connections.

Moreover, the receiving time of command in each bot can be obtained by collecting the beginning time of each bot attack in the victim or the connecting time of each botnet to a certain server for updating after receiving the update command. In some structured P2P Botnet, the approximate receiving time of command or control packet is obtained through distributed hash table (DHT) traffic monitoring [3].

The contribution of our method is to propose a reconstruction method for P2P botnet C&C channel based on the partial information that can be gathered in many types of botnets before analyzing the malware. Our method detects the edges of the network with high accuracy and its results can also be used to defend against botnet. This method is general and is not related to a particular botnet. Moreover, we use the partial gatherable information in P2P botnet, but the information used in other methods is difficult to collect in practical environments or must be complete.

In our proposed method, first, we obtain the end-to-end delay distribution on the Internet and then by using this distribution we estimate the level of each node in all cascades. The probability of each edge is calculated by considering the fact that each edge can only be linked to two nodes with the consecutive level in a cascade. Finally, the reconstructed graph is obtained by combining the probability of each edge in all of the cascades.

This paper is organized as follows. In section 2 we describe P2P botnet's characteristics and present its data gathering methods and address some forensic issues. Network reconstruction methods and their uses are studied in section 3. In section 4, we present our method and describe the steps followed. The introduced method is evaluated and the results of the simulations are discussed in section 5, and finally, our conclusion is given in section 6.

## 2. P2P Botnet

In 2003, Slapper was known as the first worm that used P2P protocol. Furthermore, in the same year, Sinit P2P botnet was discovered which used random scanning to find other peers [4]. Other P2P botnets such as Phatbot, Nugach, Storm, and Peacomm were gradually detected in the following years. Some famous P2P botnets such as Mega-D, Conficker, and TDL-4 have been propagated in recent years and their members have been estimated from 500 thousand to 10 million. It is worth mentioning that P2P botnets have a lot of members, due to limited countermeasure mechanisms against them.

Compared to centralized botnets, decentralized botnets have higher flexibility and robustness in handling a large number of bots. Since they do not have any specific point of failure, they are very hard to break down [5]. It is almost impossible to disable their C&C channels even by detecting and mitigating some of the bots.

Most decentralized botnets utilize different types of P2P protocols to construct an overlay network. This P2P overlay Network is classified as follows [2]:

- Unconstructed P2P overlay which uses alternative methods such as flooding, as it does not have any key look-up and routing features.
- Structured P2P overlay which uses routing method such as the distributed hash table.
- Super-peer overlay which has some peers with valid IP addresses as temporary servers controlling the network.

P2P networks are modeled by different types of statistical network models [6, 7] such as Erdos-Renyi random graphs, Barabasi-Albert Scale-free graph or Watts-Strogatz Small-world graph models. In the random graphs model, each edge occurs independently with equal probability. Networks following a scale-free graph model have a few numbers of nodes with high degree and a large number of nodes with low degree. In the small-world graph model, the distance of each node to most of the other nodes is smaller than a certain number of hops [8].

The following four models have been presented in [7] for botnet network models: random, scale-free, small-world, and P2P. Moreover, it should be noted that the structured P2P botnet and the unstructured P2P botnet are believed to be respectively similar to the random and the scale-free network models [7]. Super-peer botnets are similar to centralized botnets and super-peer is a single point of failure for the botnet network. So, botnets do not tend toward this design [2, 9].

Efficiency and performance of C&C channel of botnet depend on the inherent characteristics of basic graph models. These graph models are used to analyze the infection distribution in a network and its resistance against the failures of edges and nodes [10].

Studies show that the random graph model, compared to other models, is more resistant against the intentional removal of a certain node. Moreover, it is clearly observed that the removal of high-degree nodes in free-scale graphs strongly affect the connectivity of the graphs [6, 11]. Hence, we are more motivated to work on the random graph model. The focus of this paper is on the reconstruction of C&C channels for random network type of P2P botnet.

In structured P2P botnets, each node connects to at most a certain number of the peer, where this number is related to the routing mechanism in P2P protocol. Since structured P2P botnets have fixed maximum degree nodes, they are considered as random network model [7]. If unstructured P2P botnets have power-law degree distribution, then researchers use the scale-free model for their analysis. As in most of the unstructured botnets, the connections of each bot are usually limited to a certain number, and even the number of peers in a peer-list is fixed [12], so they are random networks.

In random graph $G = (V, E)$ with $N$ vertices, each node connects to other $N - 1$ nodes with the same probability. If the probability of connection between two nodes is $p$, then the probability of a node with degree $\bar{d}$ can be presented by the following binomial distribution:

$$\Pr(\bar{d}) = \binom{N-1}{k} p^{\bar{d}} (1-p)^{N-1-\bar{d}}$$

The occurrence of a big number of connections in a host is a rare phenomenon even in P2P applications. Therefore, botmaster considers a small $\bar{d}$ in order to prevent the detection of host infection. Moreover, due to the improved botnet performance and the need to prevent the disclosure of botmaster members' data, the number of nodes in a peer list is limited [13]. Botnets whose members use random scanning method to find peer nodes also are random network.

By examining some infected hosts, the botnet characteristics can be discovered in order to recognize its network model. Using honeypot techniques is one of the ways to catch an infected node. Other methods such as network behavior monitoring and reverse engineering can also help with the recognition of a botnet model. Hence, finding the value of $\bar{d}$ in botnets can be made possible through discovering the hardcoded limitation value or the number of concurrent connections with other bots.

**Table 1** Comparison of P2P botnet topology discovery methods

| Method | Mode | Required Knowledge | Behind NAT Node | Recall | FPR |
|---|---|---|---|---|---|
| Crawler [14, 17] | active | high | undetectable | medium | low |
| Sensor Node [14, 18] | passive | high | detectable | high | medium |
| Our method | passive | low | detectable | high | medium |

There are various methods for the enumeration of botnet members [14, 15]. The enumeration of centralized botnets, in which members can be detected by spying on C&C servers, is easier than that of decentralized botnets. In P2P botnets which send the members list to each other or possess super peers that have access to the list of members, botnet members can be counted by obtaining this list.

One of the enumeration methods is the examination of gathered evidence of the botnet attack. Victim's firewall log can serve as good evidence for the enumeration of botnet members. Also in P2P botnets in which each bot constantly listens to a certain port for connecting other members, the number of botnet members can be obtained by counting the addresses on the Internet with these characteristics. There are also some methods that estimate the number of botnet's nodes, using a local measurement [15].

Enumeration or estimation methods can be used to find parameter $N$ of the random graph model in which, after obtaining degree $\bar{d}$ for each node, $p$ can be found through the following equation:

$$p = \frac{\bar{d}}{N-1}$$

In our method, in addition to the parameters of the random network model, the receiving time of cascades propagated in a botnet and the end-to-end delay distribution on Internet need to be obtained. The first receiving time of cascades distributed by botnet members can be obtained through different methods such as connection log, botnet monitoring or attack evidence examining. Our method to obtain the delay distribution on the Internet is described in section 4.

## 3. Related work

For the discovering of P2P botnet topology, there has not been any method yet that can be used for all botnets and does not depend on the type of botnet. The methods that can discover the topology of P2P botnet on overlay level need to completely know the botnet and its protocol. Gaining complete knowledge of the protocol to send a message or add node requires for a thorough analysis of the botnet. This task gets even more difficult in case countermeasure mechanisms have been applied by the botnet.

In order to reconstruct scenarios of centralized botnet attacks, it is reported in [16] how the IP-level topology of the infected nodes was formed in Testbed@TWISC network, where only a limited number of nodes was infected with botnets. This method reconstructs IP-level topology of two centralized botnets by capturing all the traffics.

In the methods of crawling [14, 17] and adding sensor nodes [14, 18] for collecting data have been discussed. In these methods, if the connection information or peer-list of each node can be collected, then the graph can also be discovered. Crawling method collects the connections of the whole network by using graph search. The botnet network can also be discovered by adding sensor nodes to the botnet and gathering the connection information from each node.

In order to apply the crawling and the sensor nodes methods, we must be completely aware of the botnet protocol. The botnet must also match with the required characteristics of the methods. Some methods [19] have been presented for improving P2P botnet to oppose these discovery methods that make it impossible to send a message to peer-list from an unknown source and to add the sensor node.

Table 1 shows the properties of different methods for the discovery of the P2P botnet topology and our reconstruction method. The crawler is an active method and requires to send data in the network. So, some botnets can also bypass the crawler [20]. For applying crawler and sensor node methods, knowledge of the botnet architecture and its protocol is needed. Also, it is very time-consuming to know a botnet completely. By using the least possible information about the botnet functionality, our method can reconstruct the botnet by only monitoring the external behavior of botnet. As opposed to other methods, in the crawler method, it is not possible to discover the connections behind NAT and nodes do not have a valid IP.

The crawler and the sensor node methods have been proposed to gather information from P2P botnets, not to discover the topologies. Hence, the accuracy of the discovered topology has not been discussed. Moreover, the topology of the real botnet is unknown and so Recall (detection rate) and FPR cannot be calculated. Since the number of cascades affects our results, it is not feasible to compare the results of our method with the results of other methods. We only compare Recall and FPR of the methods in table 1 qualitatively.

Due to the lack of knowledge about the connections behind NAT, Recall is smaller in the crawler method compared to other methods, but the discovered edges have lower FPR. In the node sensor method, if the node behind NAT is not connected to the sensor for collecting the data, it will not be possible to detect the connection. While in our method, by the increase of the cascades, higher Recall and lower FPR are obtained.

There exist various fields of research, such as physics, engineering, biology, and medicine [14] with the purpose of finding the connections among the members of a certain set, and with approaches that are only different in terms of necessary data, limitations, and applications but can be considered similar in finding the edges of a graph [21].

Some methods reconstruct the graph by using neighborhood information. One of such studies is [22], in which the graph has been reconstructed by having the number of common neighbors of each node pair in a social network. But due to the lack of a mechanism for monitoring botnets' neighbors, the neighborhood information cannot be collected for P2P botnets, and the method is not applicable for P2P botnets.

For reconstructing P2P botnets, it is difficult to bypass a method that can passively collect the required data from the network and reconstruct it with high accuracy. The only data that may be collected is the time of bot reactions to the received commands. The received time of commands by each node is the base of our method for the reconstruction of the botnet topology regardless of the type of the botnet. In other words, each command in the network is distributed as a cascade, and each node reacts to the command after receiving the cascade. In this section, we review the previous studies which have used nodes receiving times of cascades for general network reconstruction.

One of the reconstruction methods [23], has sought to reconstruct the graph by using the receiving time of the cascade by each node. In this method, the graph model has not been considered, and the network reconstruction can be achieved only by using the gathered times. Finally, the most probable topology of the graph which is more compatible with the gathered times will be presented as a result.

The computational complexity of this method [23] is in super exponential order and the problem of finding the most probable graph is an NP-hard problem. The simplifications, presented to reduce the computational complexity and perform the algorithm, have led to an error increase in results. Notwithstanding, this method cannot be used in networks with many nodes; moreover, it is specific to directed graphs, while the C&C channel of P2P botnets has an undirected graph topology.

If it is possible to control some of the nodes, the introduced method in [24] can reconstruct the routing graph of the Internet, by using the partial information gathered from the "Traceroute" command to probe the paths between the nodes. Since there is no similar command in the overlay layer, this method is not applicable to overlay networks of P2P botnets.

There are some methods that reconstruct the graph through sparse recovery techniques [25, 26]. According to such methods, the graph can be interpreted as a sparse signal measured through cascades. However, the assumptions of these methods, i.e., directed edge, node degree, and weight of edges, are incompatible with P2P botnets and impractical for the reconstruction.

Another study with a probabilistic approach for the reconstruction of neural network uses the Bayesian method in the non-parametric algorithm [27]. The Cascades in neural networks are electric signals diffused in neural networks. This method reconstructs the directed network by using receiving time of cascades.

Since examining all possible graph topologies is a very time-consuming job, in this method [27], it has been assumed that each node can only receive data from a certain number of preceding nodes. Also in each step, the probability of edges is updated based on the receiving time of next cascades; finally from all probable topologies, the one with the highest probability is selected.

One of the disadvantages of methods which use Bayesian is that last cascades gain more importance and it is even likely that initial cascades get neglected. Moreover, the error in reconstruction rises with intense noise in last cascades. Another issue in the mentioned method is assigning the initial probabilities of edges which have been assumed equal in this study.

The method in [28] with a non-probabilistic approach, presents an algorithm called Netcover, which reconstructs the network by using the order of the nodes in each cascade. This method has been presented for directed graphs and is based on the rule that each node receives the data solely from its preceding nodes in the cascade.

In this algorithm [28], the input edges for each node are selected in such a way that with the least edges, all cascades conform to this rule. This method changes the issue from the reconstruction of the graph in each node to the coverage of sets and solves the problem with the greedy approach. The purpose of this greedy method is selecting the least possible number of edges and might not be reasonable in many of the applications.

Another method [29], put forward for random graphs collaboratively works with some of the participant nodes to reconstruct network. This method uses the end to end delay of four nodes for finding the topology of their middle nodes. Random graph model examined in this study reduces the number of possible topologies in the four nodes and consequently decreases the error rate in this method.

One of the disadvantages of this method [29] is that the minimal representative graph in this method is reconstructed from the main graph and might leave many nodes unrecognized. Besides, it is sometimes difficult to choose the right topology from among several possible topologies. The number of participant nodes required in this method is another disadvantage which is very hard to obtain in some practical applications.

Some of the issues that make the above-mentioned methods inefficient in the reconstruction of the C&C channel and P2P botnets include high time complexity for the big number of nodes and less accurate results. In addition, in these methods network characteristics and their applications have not been sufficiently taken into account.

Due to the disadvantages of other methods, and also the fact that reconstruction graph from temporal data of its nodes is still a challenge [30,31], in this study, we present a method which uses network characteristics and times gathered from cascades propagation to reconstruct the C&C channels of P2P botnets. If it is possible to select the root node for the cascade propagation, future cascades can be propagated from a node with more influence on the correction of the then obtained results.

## 4. Our method

We have adopted a probabilistic approach in our method and used partial gatherable data from P2P botnet networks. This data can be gathered from both directions of connections between bots. In our method, like most network reconstruction methods, the graph nodes are known. Attacked target logs or revealed peer lists constitute a good source for finding graph nodes and their IP addresses.

After collecting the necessary data by observing some cascades, our network reconstruction method is done in three steps. In the first step, the level of each observed node in each cascade is estimated, and in the second step, nodes level distribution in cascades is extracted. In the third step, the more probable edges are detected and presented as a reconstructed network.

### 4.1. Level Discovery

The level of a node refers to the smallest hop count between cascade root and the node. In the first step, the level of each node in a cascade is determined based on the receiving time of each node and the delay distribution between each node pair in the overlay network. In our method, node level is defined as the shortest hop count distance between node and cascade root.

In our method, we assume to have a random graph model and the delay distribution of overlay connection between two peers. Random graph model parameter $N$ is equal to the number of botnet members whose enumeration method has been explained before. Also, parameter $p$ of this model is obtained from node degree distribution by observing the connection degree of some bots or reverse engineering of botnet protocol.

In [32], it has been concluded that the delay between any internet node pair has a Gamma distribution, so the delay distribution between two botnet nodes is described as

$$Pr(\tau) = \frac{1}{\Gamma(k)\theta^k} \tau^{k-1} e^{-\frac{\tau}{\theta}} \quad (1)$$

where $k$ and $\theta$ are shape and scale parameters of Gamma distribution and $\tau$ is delay. Due to the delay restriction on data transmission between nodes, a time constraint is defined on the delay between nodes $a_i$ and $a_j$. That is the delay of each node is limited between $\tau_{min}$ and $\tau_{max}$ and the delay distribution is truncated as:

$$P_{delay}(\tau) \triangleq Pr(\tau | \tau_{min} < \tau \leq \tau_{max})$$
$$= \frac{\frac{1}{\Gamma(k)\theta^k} \tau^{k-1} e^{-\frac{\tau}{\theta}}}{F(\tau_{max}) - F(\tau_{min})} \quad (2)$$

where $F(\tau) = \frac{1}{\Gamma(k)} \gamma(k, \frac{\tau}{\theta})$ is a cumulative distribution function for delay $\tau$. If, after collecting the receiving time in each node for a cascade, the time difference of that node pair is less than $\tau_{min}$ or more than $\tau_{max}$, then the edge probability is zero.

After obtaining the truncated delay distribution and collecting the receiving times of all cascades, the following sequence is generated for each cascade $c$ by sorting the nodes based on receiving time.

$$<a_1^c, t_1^c>, <a_2^c, t_2^c>, \ldots, <a_N^c, t_N^c>: \forall i < N \; t_i^c \leq t_{i+1}^c$$

If the difference between the receiving time of a node and its immediate previous node is bigger than $\tau_{max}$, or the difference between receiving time of the node and all preceding nodes is smaller than $\tau_{min}$, or there is no preceding node, then the node is considered to be in level one. So, the probability of the first level for this node is equal to one, and the probability of level $L$ for other nodes is obtained using the following equation 3.

$$P_{level}(l_k^c = L)$$
$$= \frac{\sum_{i=1}^{k-1} P_{level}(l_i^c = L-1) P_{delay}(t_k^c - t_i^c)}{\sum_{j=1}^{N}(\sum_{i=1}^{k-1} P_{level}(l_i^c = j-1) P_{delay}(t_k^c - t_i^c))} \quad (3)$$

where $l_k^c$ denotes the level of node $a_k$ in cascade c and $t_k^c$ is the receiving time of cascade $c$ for node $a_k$. After calculating the probability of any node of cascade $c$ being in level $L$, the hop distance between that node and the cascade root is equal to the highest probability level.

### 4.2. Level Distribution

In [33], it has been shown that the shortest path in random networks has a Weibull distribution. Hence, level Distribution for nodes follows the same Weibull distribution, as level distribution and shortest path are essentially similar problems. In our simulation of random graphs, we observed that nodes level distribution of all cascades in these networks is very close to the shortest path distribution and follows Weibull distribution, as the experiments described in section 5 indicates the same conclusion.

So, by using the parameters of the random network model of C&C channel, we can find the distribution of nodes level in all cascade in this network, and consequently, estimate the number of nodes in each level. Moreover, in each level, we can find the number of hidden nodes whose receiving time of cascade could not be gathered.

In the random networks, the smaller the probability $p$, i.e. the fewer nodes connection, the bigger the domain of distribution and the graph diameter. The obtained distribution can be used to estimate the number of invisible nodes of each level and even levels in which no nodes have been observed.

In order to estimate the number of nodes in each level, we must obtain Weibull distribution of nodes level in the graph. Hence, first we generate some instances of the graph on the basis of the random network model parameters $N$, and $p$, then we find the levels of nodes in all possible cascades. Finally, after obtaining the histogram of levels, we fit Weibull distribution to this histogram.

The occurrence of a case, where there is a gap between estimated levels, i.e., where a node level is estimated to be 1, despite the existence of previous nodes in the sequence, shows the existence of some levels of which no member nodes have been observed. In this case, the hop distance between the nodes that immediately precede the gap and the first nodes is definitely more than one and there is no edge between these two sets of nodes.

Although solving the challenge of the existing gap size does not affect our network reconstruction method, it is necessary to estimate the number of invisible nodes in the gap for determining the existence of any unknown nodes at the beginning or the end of the sequence. To achieve this, first, we calculate the centroid $T_l^c$ of each level. A centroid can be calculated through the following equation:

$$T_l^c = \frac{\sum_{i=1}^{N} t_i^c P(l_k^c = l)}{\sum_{i=1}^{N} P(l_k^c = l)} \quad (4)$$

$$\Delta T_{AVG}^c = \frac{\sum_{i=2}^{maxLevel} T_i^c - T_{i-1}^c \quad (if\ in\ level\ i, i-1\ has\ at\ least\ one\ member)}{number\ of\ consecutive\ centroids} \quad (5)$$

$$w_{e_{ij}}^c = \sum_{L=1}^{N-1} P_{level}(l_i^c = L) P_{level}(l_j^c = L+1) P_{delay}(t_j^c - t_i^c) + \sum_{L=2}^{N} P_{level}(l_i^c = L) P_{level}(l_j^c = L-1) P_{delay}(t_i^c - t_j^c) \quad (6)$$

Then, we estimate the time difference average for consecutive centroids by equation 5. After that, we estimate the number of invisible levels between the two node sets. This is equal to the number of $\Delta T_{AVG}^c$ located between the centroid time of the preceding and the following levels of the gap.

After finding the number of middle gaps levels, the estimated levels are modified. In order to do this, starting from the beginning of the sequence, the correction value is obtained by adding the number of estimated levels of each gap and the last preceding level. The correction value is then added to all the following levels in the sequence.

If the biggest estimated level is equal to the diameter of the graph obtained from Weibull distribution, the probability of gap in the beginning or the end of the sequence is very low. Otherwise, we find the closest distribution to Weibull by increasing the estimated level of all nodes several times if necessary. In this way, the number of beginning and ending invisible levels are obtained. One of the advantages of detecting invisible levels is finding the distance between the origin of command and the visible nodes or the origin itself in P2P botnets.

### 4.3. Network Reconstruction

The method we use for network reconstruction is based on the fact that there is no edge between two nodes whose level difference is 2 or more in at least one cascade. If there is an edge between two nodes, then their estimated level difference must be less than or equal to 1, otherwise, there is definitely no edge between them. In other words, if by considering the delay distribution, the delay probability for the time difference of each node pair is zero, then there are no edges between them.

Thus, in our method after estimating the level of each node in cascade c, we add edge $e_{ij}$ with weight $w_{e_{ij}}^c$ to temporary edges, to indicate our confidence on the existence of the edge between nodes $a_i$ and $a_j$. The weight $w_{e_{ij}}^c$ is calculated by equation 6 for two nodes of a link with estimated successive levels.

We calculate the weight of each edge in different cascades and if in at least one cascade this weight becomes zero, our assumption about the existence of this edge is contradicted. In this case, we will remove this edge from temporary edges.

After observing the receiving time of each cascade and calculating its edge weight, the total of the edge weight $W_{e_{ij}}$ for all the observed cascades is calculated by equation 7. The reconstructed edges are selected from the temporary edges with the biggest total weight.

$$W_{e_{ij}} = \sum_{c=1}^{C} w_{e_{ij}}^c + \sum_{c=1}^{C} w_{e_{ji}}^c \quad (7)$$

Algorithm 1 provides the pseudocode of our method to reconstruct the graph by having the receiving times of $M$ cascades. The time complexity of this algorithm is $O(maxLevel \times MN^2)$, where $N$ is the number of nodes, and $maxLevel$ is the maximum level of each node in all cascades, that is calculated for certain random graph. In this algorithm, first, the level probability of all peers for each cascade is obtained, and then the edge probability is calculated. Finally, the edge probabilities are combined for all cascades, and the reconstructed graph is returned.

It should be noted that the number of the selected reconstructed edges is equal to the mean of the random graph edges, i.e., $p(N(N-1)/2)$. Moreover, the reconstructed edges are selected from the temporary edges having a delay probability of more than zero in all cascades.

**Algorithm 1** our reconstruction method

**Input:** an $N$-member list of peers, receiving times for $M$ cascades, link probability $p$
**Output:** a reconstructed graph

for each edge $e_{ij}$, where $0 < i, j \leq N$:
   $W_{e_{ij}} \leftarrow 0$
*temporary-edges* $\leftarrow$ *(all of the possible edges)*
for each cascade $c_m$, where $0 < m \leq M$:
   *sorted-node-list* $\leftarrow$ *sort-by-time*(nodes of $c_m$)
   for each node $k$ in *sorted-node-list*:
     for each level $L$, where $0 < L \leq maxLevel$:
       Calculate $P_{level}(l_k^{c_m} = L)$ by equation (3)
   for each edge $e_{ij}$, where $0 < i, j \leq N$:
     for each level $L$, where $0 < L \leq maxLevel$:
       Calculate $w_{e_{ij}}^{c_m}$ by equation (6)
     if ($w_{e_{ij}}^{c_m} = 0$) then
       remove $e_{ij}$ from *temporary-edges*
   $W_{e_{ij}} \leftarrow w_{e_{ij}}^{c_m} + w_{e_{ji}}^{c_m} + W_{e_{ij}}$
*edge-list* $\leftarrow$ *sort-by-W*(*temporary-edges*)
*reconsGraph* $\leftarrow$ select $(\frac{N(N-1)}{2} \times p)$ edges from *edge-list*
return *reconsGraph*

## 5. Evaluation

In order to evaluate our method in this paper, we use Recall and false positive rate (FPR) metrics. Recall is the fraction of the correctly reconstructed edges of the actual graph. FPR is the fraction of edges that do not exist in the actual graph but are incorrectly reconstructed. Using these metrics, we assess the accuracy of our method with respect to the similarity of the reconstructed graph to the actual one.

In our method, the number of reconstructed edges for any random graph with a definite number of nodes is chosen to be equal to the average number of the edges of that graph. Then, the number of incorrectly reconstructed edges is approximately equal to the number of edges in the real graph that is missed in our results. So, the value of Precision metric is very close to Recall. In this paper, we do not deal with Precision metric separately.

Since it is difficult to know the topology of real P2P botnets to compare the reconstructed graph to the actual one, and also there exists no data set for this purpose, we use simulation to evaluate our method. Another reason for using simulation is that it is impractical to generate the C&C channel topology by providing a set with a large number of supervised systems infected by the bot and dispersing them in different locations.

The C&C channel of a P2P botnet is represented by a graph. In simulating Erdos-Renyi random graph for structured P2P botnets, the number of nodes and the probability of each edge between node pair should be known. First, all $N$ nodes are labelled, then an edge is assumed between every possible node pair with p probability. After that, a delay value is assigned for the assumed edges based on the end to end delay distribution on internet path. According to these delays, the data is propagated throughout the simulated C&C channel and received by all nodes.

For simulating cascades, one node out of $N$ nodes is selected as the source node from which the data is propagated throughout the graph. The data is received by each node after being forwarded by adjacent nodes and later than the edge delay. In this simulation, we assume no delay in the nodes and each node immediately forwards a data to all adjacent nodes after receiving it for the first time but does not show any reaction when receiving it in the next times

The first receiving times of data by nodes are inputs of our reconstruction method. In our evaluation, the receiving time of data by each node is indicated by observing the reaction of the node to the data. For this reason, the collected times include the delay between the node and the observer. Since this time is due to the end to end delay in the internet, we assigned its value by using the delay distribution of graph edges in simulation. In other words, the difference between the observed time and the real receiving time is as much as one graph edge delay.

Since in each simulation, both the formation of the graph and the order of the cascades are random, we run each simulation 10 times to obtain Recall and FPR by calculating the average of their values in all runs.

### 5.1. Delay Distribution

In order to find the delay distribution of an edge of a real botnet on the Internet, we can use the precomputed delay distributions. These distributions are calculated by considering the location of the nodes related to the edge. In our simulation, the delay distribution of all edges is assigned based on a single delay distribution, because there are no precomputed delay distributions for any location of the nodes.

In order to estimate the delay distribution, we used the data collected in NCC RIPE TTM project and fitted the histogram of internet delay between Middle East Technical University (METU) and Gebze Institute of Technology (GIT) to Gamma distribution in order to obtain its parameters. We obtained Gamma distribution with shape parameter k=42.27 and scale parameter θ=0.35. Also, the minimum and maximum delay between METU and GIT are 5 and 517 milliseconds.

### 5.2. Level Discovery

After collecting the receiving times in a cascade, we estimate node level using equation 3. In order to evaluate the first step of the method, we obtain Recall for node level. Recall for node level is the ratio of the number of correctly estimated node level to the total number of nodes in each cascade.

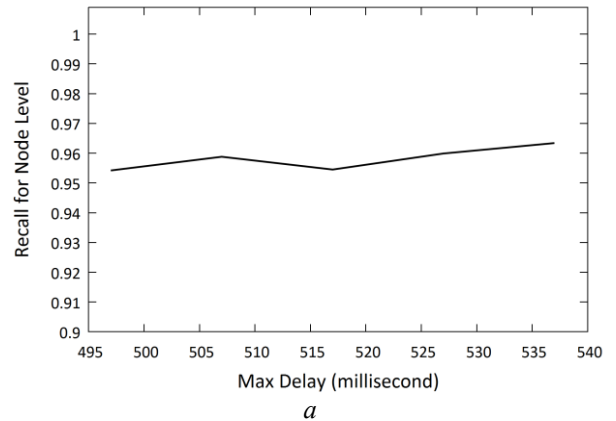

*a*

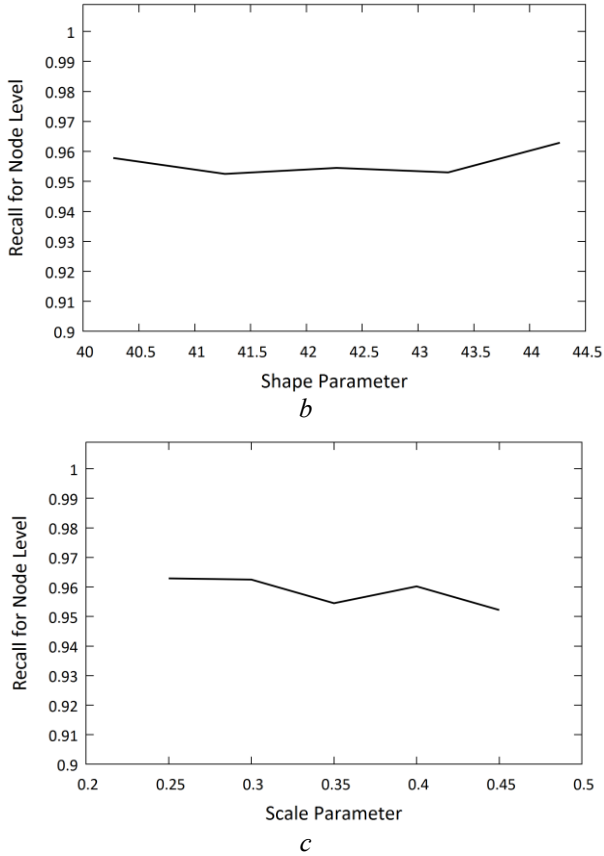

***Fig. 1***. *Recall for Node Level vs. Delay Distribution Parameter Changes*
*(a) Max Delay change, (b) shape change, (c) scale change*

Since the end-to-end delay distribution fitted in the previous section may have error and its parameters are not accurate, the changes of Recall for node level by each parameter of the delay distribution are examined. As shown in figure 1, Recall is hardly affected even with error in the estimation of the delay distribution parameters. In our study, this rate has been more than 89% in all the studied cases.

Recall for node level obtained for all cascades of the graphs with 1000 nodes and the edge probabilities 0.01, 0.02, 0.03, …, 0.10 are illustrated in Figure 2. The accurately detected node levels will be more than %92. With the increase of link probability, Recall for node level also gets bigger. This observation suggests that the bigger the link probability, the smaller the graph diameter and the less the error rate in level estimation. However, graphs with bigger diameters have more levels, so the accuracy of their estimated level is smaller.

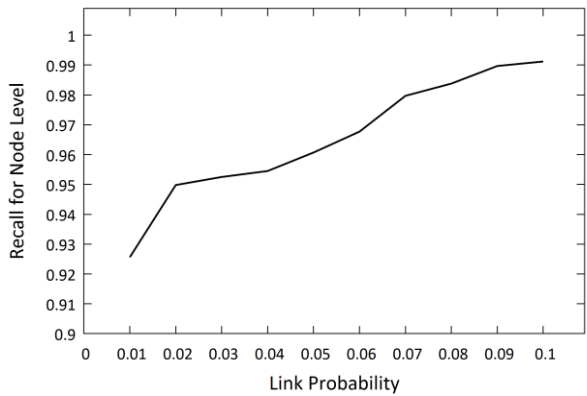

***Fig. 2***. *Recall for Node Level vs. Various Link Probabilities*

The node level estimated with respect to the edge delay and the receiving time, might be different from the node level calculated with respect to the hop count. As a cascade propagation in a graph is reconstructed in a tree format, such a difference will not affect the results of reconstruction. In our method, the sequence of node level is used for reconstructing the edge and their exact levels are not important.

### 5.3. Level Distribution

In this simulation, in order to obtain the distribution for levels of nodes in cascades, first, each node is once considered as the root from which a cascade is propagated. Then the level distribution is obtained for that cascade. Finally, in order to aggregate the results, the sum number of nodes in each cascade is calculated and the level distribution for all cascades of the graph is obtained.

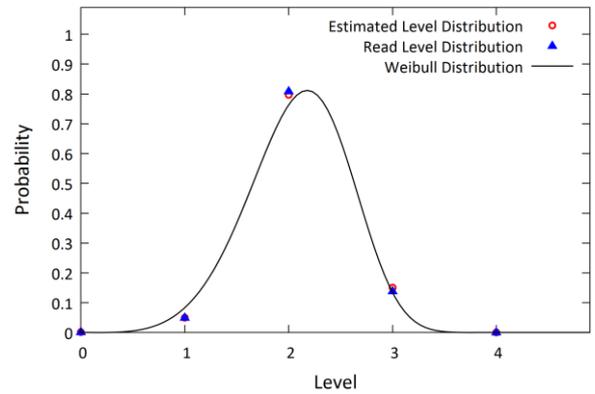

***Fig. 3***. *Weibull Distribution fit to Level Distribution*

Our examinations on N nodes with different configuration, show that these two distributions are similar, and the number of nodes in each level can be estimated by fitting it to Weibull distribution. As shown in Figure 3, Weibull distribution with shape parameter $k=4.923$ and scale parameter $\theta=2.281$ is obtained for a graph with 1000 nodes and link probability 0.05.

The number of nodes in each level can be estimated by obtaining the Weibull distribution of node level in a random graph. If all receiving times in a cascade cannot be collected, the number of invisible nodes in each level can be estimated by using Weibull distribution. However, recognizing the invisible nodes in each level is not dealt with in this paper. Our method reconstructs the graph by using merely the level of the visible nodes.

### 5.4. Network Reconstruction

In order to evaluate the accuracy of our reconstruction method, after propagating each cascade of a graph, the weight of all possible edges are calculated by using equation 6. Then, the edges with more weight are presented as the reconstructed graph. Recall curve for the edge, in a graph with 1000 nodes and a link probability of 0.05 has been shown in Figure 4. As it can be seen in this figure, Recall for edge rises with the increase in the number of cascades, and Recall of 90 is obtained with 22 cascades.

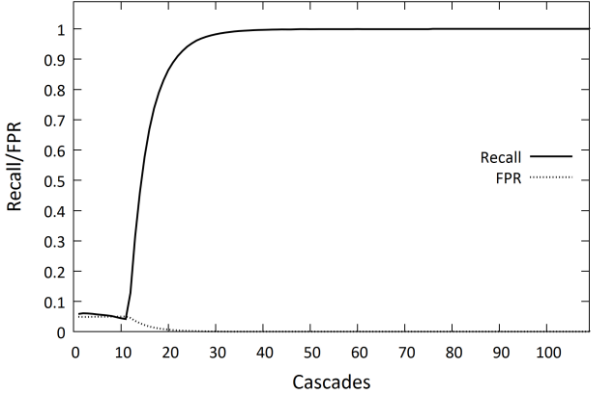

*Fig. 4. Recall vs. various number of cascade*

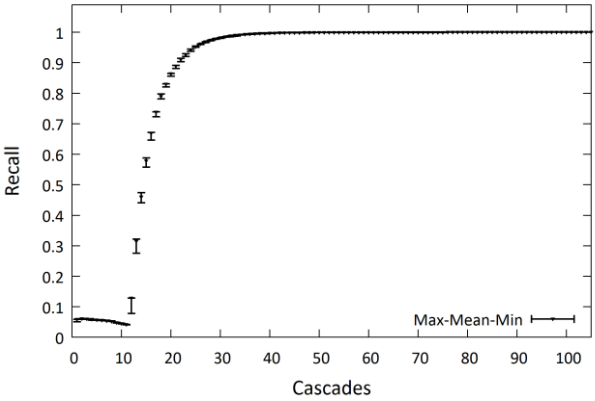

*Fig. 5. Recall (Min, Mean, Max)*

In Figure 5, in addition to mean value, the minimum, and maximum difference have been illustrated to show the variance of the obtained Recall for each cascade. For example, with 14 cascades in a network with 1000 nodes and node degree mean 50, min, max, and average mean values are respectively 0.4407, 0.4739, and 0.4599. From Figure 5, no noticeable difference is observed among max, min and average Recall in different examinations.

### 5.5. Required Cascades

One of the main questions regarding our method is how to calculate the number of required cascades in network reconstruction to obtain a suitable Recall value in real P2P botnets. The number of all distinct cascades that can be propagated in a network is equal to the number of nodes in that network. In other words, the root of each cascade can be any node in the network resulting in a different cascade.

As the number of members for any real P2P botnet is different, in order to assess the efficiency of our method for different types of botnets, the graphs have been selected with the different number of nodes in our experiments. Also, using different P2P protocols for implementation of C&C channel, results in the variation of node degrees and in other words, in our simulations the edge probability in random graph model is different.

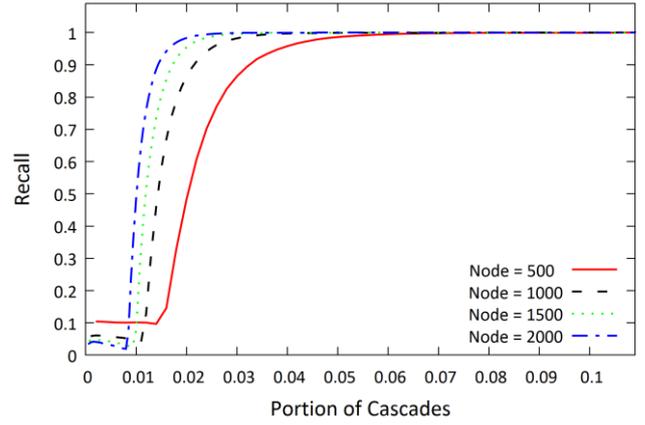

*Fig. 6. Recall for edge vs. number of graph nodes*

Thus, in order to obtain the number of required cascades for the reconstruction of the whole network, four graphs are examined. In these examinations of the random networks - with the same node degree mean distribution and the different number of nodes - Recalls are obtained per cascade added, until all cascades are covered in our method. Figure 6 shows Recalls of network reconstruction per cascades proportion increased. These networks have 500, 1000, 1500, and 2000 nodes and the node degree mean 50.

In these examinations, it was observed when the number of network nodes increases, we may obtain a certain Recall with a smaller proportion of all cascades. For example, to obtain a Recall of 90% for networks with 500, 1000, 1500, and 2000 nodes, respectively 0.034, 0.022, 0.017, and 0.014 of all cascades are needed.

If the mean of a node degree distribution is known, the probability of the edge in a random graph model is equal to the ratio of this mean to the total number of nodes. As in different P2P protocols, the number of adjacent nodes of each bot is limited, we do our experiments with different node degrees using the edge probability of random graph model obtained from those node degrees.

In order to examine the effect of node degree on our method, Recalls for three graphs with 1000 nodes and the degree mean of 10, 20, 30, … 100 are obtained as shown in Figure 7. With our method, Recall gets smaller with the decrease in node degree and the increase in graph sparseness. Despite this, with 100 cascades out of 1000, Recall is still bigger than 0.92.

According to the obtained results, our method is evaluated in even the worst case where the mean of graph degree is so small that the graph does not have a cycle. Although Recall of node level for a 1000-member graph is 0.27, by using 315 cascades, the edge Recall reached 0.90.

Given the same node degree, it can be deduced that the decrease in the proportion of cascades needed to obtain a certain Recall, is due to the sparseness of graphs. The sparser the graph, the smaller the number of required cascades for the reconstruction of an adequate proportion of edges of the network. As a result, for real P2P botnets that have a great number of nodes, with even smaller cascade proportion, more accurate results are reached.

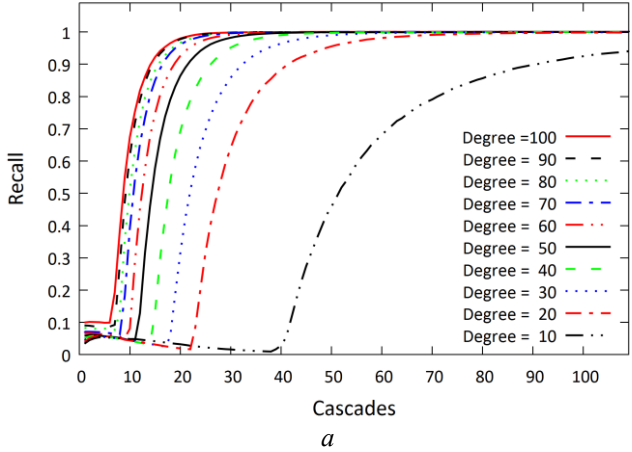

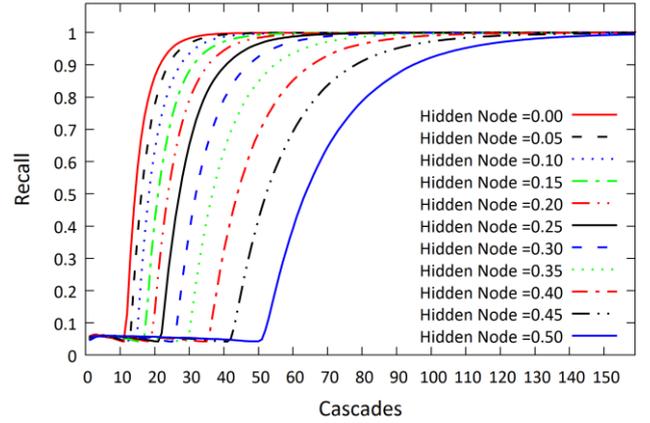

*Fig. 8. Recall vs. Hidden Nodes Percent*

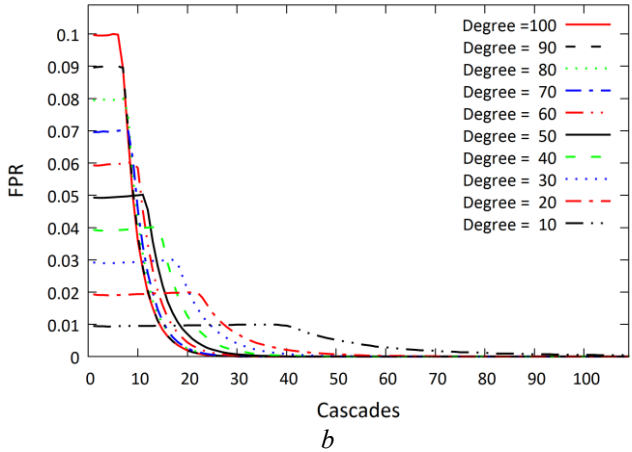

*Fig. 7. Recall and FPR vs. Node Degree changes (a) Recall, (b) FPR*

### 5.6. Partial Observations

The results of our method have been examined in this section, assuming that it is impossible to obtain the receiving time of cascades by all the nodes. In order to do the examinations, we hide some of the data and exclude them from our observations in the simulation. This portion of data actually belongs to the part of nodes for which no data has been collected. These nodes are selected randomly from each cascade.

The results of our method have been shown in Figure 8 for %10, %15, %20, …, %50 of the invisible nodes. The more the share of the invisible nodes, the more cascades we need for obtaining the threshold. But the pattern of Recall increase is somehow similar in all observations. We can even reach Recall 0.90 with only half of the data from a 1000-node graph with node degree mean 50.

## 6. Conclusion

In this paper, we have proposed a method for the reconstruction of P2P botnet C&C channel to help to resolve the problems and ambiguities in the investigation of P2P botnets. Considering the limitation of P2P botnet, our method reconstructs the C&C topology by collecting the receiving times of a cascade by each node and by using a random model network of C&C channel.

In this method, at first, we obtain the probability of each node in each level of a cascade. Then based on that, the probability of edge in that cascade is obtained. Then by repeating this method for other cascades, we try to make the edge probability closer to the actual value. Finally, we propose the edges with the highest probability as the network edges.

Our observation shows that this method can estimate the C&C channel with a good Recall by considering our evaluation results. The more our knowledge of the delay distribution and random graph model, the more accurate the obtained results. Moreover, with an increase of cascade numbers, the detection rate of our method increases, and the false positive rate decreases. If it is possible to distribute the forged cascades, then it is likely to obtain a considerable number of cascades quickly.

In our method, Recall is higher for graphs with bigger node degree mean. The reason is that random graphs with lower node degree have a bigger diameter and a smaller Recall of node level. This condition results in the estimation of more wrong edges. Also, with the increase of cascades number, the distance between nodes may get bigger than one level, so the wrong results are corrected.

The advantage of our method is that the C&C channel of P2P botnet can be reconstructed with acceptable complexity and high accuracy. This method is general and can be applied to a large number of P2P botnets. The novelty of our method is using end-to-end delay on the Internet for the first time and the probabilistic approach to reconstruct the topology based on P2P botnet characteristics. However, it cannot be applied to scale-free P2P botnets and the diversity of roots in observed cascades affects the results.

Our proposed method can raise several open problems. By using the obtained topology, we can propose more efficient containment methods, and detect botmaster or its directly connected nodes. By using the reconstructed topology, the infection distribution path can be found, and the infection type can be detected. In addition, this method can be used to reconstruct other random networks such as neural network or file sharing networks.